\documentclass[fleqn,usenatbib]{mnras}

\usepackage{newtxtext,newtxmath}

\usepackage[T1]{fontenc}

\DeclareRobustCommand{\VAN}[3]{#2}
\let\VANthebibliography\thebibliography
\def\thebibliography{\DeclareRobustCommand{\VAN}[3]{##3}\VANthebibliography}

\usepackage{amsmath}	% Advanced maths commands
\usepackage{graphicx}	% Including figure files
\usepackage[caption=false]{subfig}

\newcommand{\kms}{km s$^{-1}$}

\title[Condor Observations of NGC 5907]{Introducing the Condor Array Telescope.
II. Deep imaging observations of the edge-on spiral galaxy NGC~5907 and the
NGC~5866 Group:  yet another view of the iconic stellar stream}

\author[Lanzetta, Gromoll, Shara, et al.]{
Kenneth M. Lanzetta,$^{1}$\thanks{E-mail: Kenneth.Lanzetta@stonybrook.edu}
Stefan Gromoll,$^{2}$
Michael M. Shara,$^{3}$
Stephen Berg$^{1}$,
\newauthor{James Garland,$^{3}$
Evan Mancini,$^{1}$
David Valls-Gabaud,$^{4}$
Frederick M.\ Walter,$^{1}$
}
\newauthor{and John K.\ Webb$^{5}$}
\\
$^{1}$Department of Physics and Astronomy, Stony Brook University, Stony Brook,
NY 11794-3800, USA \\
$^{2}$Amazon Web Services, 410 Terry Ave.\ N, Seattle, WA 98109, USA \\
$^{3}$Department of Astrophysics, American Museum of Natural History, Central
Park West at 79th St., New York, NY 10024-5192, USA \\
$^{4}$Observatoire de Paris, LERMA, CNRS, 61 Avenue de l'Observatoire, 75014,
France \\
$^{5}$Institute of Astronomy, University of Cambridge, Madingley Road,
Cambridge CB3 0HA, United Kingdom
}

\pubyear{2023}

\begin{document}
\label{firstpage}
\pagerange{\pageref{firstpage}--\pageref{lastpage}}
\maketitle

\begin{abstract}
We used the Condor Array Telescope to obtain deep imaging observations through
the luminance filter of the entirety of the NGC 5866 Group, including a very
extended region surrounding the galaxy NGC 5907 and its stellar stream.  We
find that the stellar stream consists of a single curved structure that
stretches $220$ kpc from a brighter eastern stream to a fainter western stream
that bends to the north and then curls back toward the galaxy.  This result
runs contrary to a previous claim of a second loop of the stellar stream but is
consistent with another previous description of the overall morphology of the
stream.  We further find that: (1) an extension of the western stream appears
to bifurcate near its apex, (2) there is an apparent gap of $\approx 6$ kpc in
the western stream due east of the galaxy, (3) contrary to a previous claim,
there is no evidence of the remnant of a progenitor galaxy within the eastern
stream, although (4) there are many other possible progenitor galaxies, (5)
there is another structure that, if it is at the distance of the galaxy,
stretches $240$ kpc and contains two very large, very low-surface-brightness
``patches'' of emission, one of which was noted previously and another of which
was not.  We not the number and variety of stellar streams in the vicinity of
NGC 5907 and the apparent gap in the western stream, which may be indicative of
a dark subhalo or satellite in the vicinity of the galaxy.
\end{abstract}


\section{Introduction}

Over the past several years, the subject of low-surface-brightness imaging of
astronomical sources has experienced a resurgence of interest, driven by new
instrumentation capable of recording low surface brightnesses over wide fields
of view.  The edge-on spiral galaxy NGC 5907 has become a prime target of such
observations.  The galaxy was discovered by William Herschel in 1788 using an
18.7-inch (47.5 cm) reflecting telescope \citep{her1789} and is a member of the
NGC 5866 Group, which consists of at least the galaxies NGC 5866 (or M102), NGC
5879, and NGC 5907.  The NGC 5866 Group is located near on the sky to the M101
Group and the M51 Group, and the redshifts of all three groups are similar,
which suggests that they are all part of the same structure.

Observations of NGC 5907 in H I by \citet{san1976} showed that the galaxy
exhibits a pronounced warp, which was also observed at optical wavelengths by
\citet{van1979}, \cite{sas1987}, and \cite{sac1994}.  Subsequent observations
of the galaxy at optical wavelengths by \citet{sha1998} and \citet{zhe1999}
revealed a remarkable stellar stream forming a section of a loop surrounding
the disk of the galaxy.  The galaxy and the stellar stream were then observed
again at optical wavelengths by \citet{mar2010}, who reported that the stellar
stream comprised not one but {\em two} full loops surrounding the disk of the
galaxy and proposed that both loops could be plausibly modeled by $N$-body
simulations as the accretion of a dwarf satellite onto the disk of the galaxy.
Because the configuration was so striking and unusual, this image of NGC 5907
by \citet{mar2010} became {\em the} iconic image depicting the effects of tidal
interactions between accreting dwarf satellites and spiral galaxies and is
likely one of the most widely-recognized and influential images of any galaxy
ever.  The galaxy was then observed again by \citet{lai2016} using the
Suprime-Cam imager on the Subaru 8.2-m telescope through the Sloan $g'$, $r'$,
and $i'$ filters and using the Infrared Array Camera on the Spitzer telescope
at 3.6 $\mu$m; these observations detected only the first loop of the stellar
stream.

The situation took another dramatic turn when \citet{van2019} used the
Dragonfly Telephoto Array \citep{abr2014} to again observe the galaxy and the
stellar stream at optical wavelengths.  These observations (1) showed no
evidence at all of the second loop of the stellar stream but instead (2)
indicated that the stellar stream consists of a single curved structure that
stretches 220 kpc, from the brighter ``eastern stream'' or the first loop
identified by \citet{sha1998} and \citet{zhe1999}, across the southern edge of
the galaxy, to a fainter ``western stream'' that bends to the north.  Results
of \citet{van2019} further indicated (3) a ``density enhancement near the
luminosity-weighted midpoint of the [eastern] stream,'' which they interpreted
as the ``likely remnant of a nearly disrupted progenitor galaxy,'' (4) that
the configuration could be plausibly modeled by $N$-body simulations, (5) a
new ``linear'' feature emanating from the eastern stream toward the east and
terminating on a ``patch'' of emission (6) a tentative extension of the western
stream to the northeast looping back south toward the disk, (7) a tentative
continuation of the eastern stream looping back to the disk, and (8) a
previously-uncataloged dwarf galaxy located just west of the eastern stream.
Subsequent observations at optical wavelengths by \citet{mul2019} and by
\citet{byu2022} likewise showed no evidence at all of the second loop, although
these observations also did not detect the western stream or any of the other
features reported by \citet{van2019}.

In the late winter and spring of 2022, we used the Condor Array Telescope
\citep{lan2023} to obtain deep imaging observations through the luminance
filter of the entirety of the NGC 5866 Group, including a very extended region
surrounding the galaxy NGC 5907 and its stellar stream.  Our motivation was
severalfold:
\begin{itemize}

\item to assess the technical capabilities and sensitivity of Condor in
comparison with other telescopes optimized for low-surface-brightness imaging,
which is especially relevant since NGC 5907 and its stellar stream have become
something of a benchmark within the low-surface-brightness community;

\item to confirm (or refute) the results of \citet{van2019} and to weigh in on
the apparent discrepancy between the results of \citet{van2019} and the results
of \citet{mar2010};

\item to search for new low-surface-brightness features in the vicinity of
NGC 5907, potentially with greater sensitivity than any previous observations;

\item to exploit the higher angular resolution of Condor with respect to
Dragonfly to help constrain the nature of the various low-surface-brightness
features in the vicinity of NGC 5907;

\item and to set low-surface-brightness features in the vicinity of NGC 5907
into the broader context of the NGC 5866 Group.

\end{itemize}
Here we report results of these observations, which together constitute the
deepest imaging observations of NGC 5907 and its stellar stream and of the NGC
5866 Group yet obtained.  In what follows, we adopt for the galaxy NGC 5907 a
heliocentric recession velocity $v = 665 \pm 1$ \kms\ and redshift $z =
0.002218 \pm 0.000002$ \citep{spr2005} and a distance $d \approx 17$ Mpc
\citep{tul2016}.

\section{Observations}

Condor is an ``array telescope'' that consists of six apochromatic refracting
telescopes of objective diameter 180 mm, each equipped with a large-format
($9576 \times 6388$ pix$^2$), very low-read-noise ($\approx 1.2$ e$^-$), very
rapid-read-time ($< 1$ s) CMOS camera.  Condor is optimized for measuring {\em
both} point sources {\em and} extended, very low-surface-brightness features
and in its normal mode of broad-band operation obtains observations of exposure
time 60 s over dwell times spanning dozens or hundreds of hours.  In this way,
Condor builds up deep images while simultaneously monitoring tens or hundreds
of thousands of point sources per field at a cadence of 60 s.  Details of the
motivation, configuration, and performance of the telescope are described by
\citet{lan2023}.

In the late winter and spring of 2022, we used Condor to obtain deep imaging
observations through the luminance filter of the entirety of the NGC 5866
Group.  The Condor images differ from the Dragonfly images of NGC 5907 in three
significant ways: (1) they are of higher angular resolution (with a plate scale
of $0.85$ arcsec pix$^{-1}$ for Condor versus $2.8$ arcsec pix$^{-1}$ for
Dragonfly), (2) they extend over a wider field of view, and (3) they were
obtained at a more rapid cadence.  Here we consider only deep images formed
from sums of the individual exposures, neglecting any temporal aspects of the
observations; we defer consideration these other aspects of the observations
until elsewhere.  These observations targeted NGC 5907 and the NGC 5866 Group
in six different pointings:  five pointings to ``Condor
fields''\footnote{``Condor fields'' are set of fields with field centers that
tile the entire sky with the Condor field of view, allowing for overlap.} and
one pointing centered on NGC 5907.  All observations were obtained with an
individual exposure time of 60 s, and the telescope was dithered by a random
offset of $\approx 15$ arcmin between each exposure.  Details of the
observations are presented in Table 1, which for each pointing lists the
International Celestial Reference System (ICRS) coordinates of the field center
and the total exposure time.  The total exposure time summed over the six
pointings is 122 hr.

\begin{table}
\centering
\caption{Details of observations.}
\begin{tabular}{p{1.05in}rrc}
\hline
\multicolumn{1}{c}{ } &
\multicolumn{2}{c}{J2000} &
\multicolumn{1}{c}{Exposure} \\
\cline{2-3}
\multicolumn{1}{c}{Pointing} & \multicolumn{1}{c}{R.A.} &
\multicolumn{1}{c}{Dec} & \multicolumn{1}{c}{(h)} \\
\hline
Condor field 6089 \dotfill & 15:05:34.03 & $+$55:54:32.76 & 22.7 \\
Condor field 6090 \dotfill & 15:20:24.74 & $+$55:54:32.76 & 25.0 \\
Condor field 6183 \dotfill & 14:58:03.86 & $+$57:16:21.72 & 19.7 \\
Condor field 6184 \dotfill & 15:13:32.90 & $+$57:16:21.72 & 24.1 \\
Condor field 6185 \dotfill & 15:29:01.94 & $+$57:16:21.72 & ~~9.1 \\
NGC 5907 \dotfill          & 15:15:53.69 & $+$56:19:43.86 & 21.4 \\
\hline
\end{tabular}
\end{table}

\section{Condor Data Pipeline Processing}

We processed the observations described in \S\ 2 through the Condor data
pipeline.  The data pipeline processing proceeds in several steps as follows:

\begin{enumerate}

\item Each science image is bias subtracted using a ``master bias'' image
determined from a sequence of 500 zero-second exposures.  The master bias image
appropriate for a particular science image is typically obtained on the morning
immediately preceding or following the acquisition of the science image.

\item The width of the central region of the autocorrelation function and the
average sky background level of each science image are measured and recorded.
These values are used subsequently to assess the quality of the science images.

\item Each science image is field flattened and background subtracted.  As
described by \citet{lan2023}, this involves dividing the science image by an
appropriate ``twilight flat image'' (i.e.\ a sum of images of the sky obtained
during dusk or dawn twilight), masking regions of the image surrounding
detectable sources using NoiseChisel \citep{akh2015, akh2019}, fitting the
resulting quotient with a high-order (typically eighth-order) two-dimensional
polynomial, and subtracting the resulting polynomial fit from the quotient.
Because the source mask depends on the background, this procedure is iterated
through convergence (which typically requires four iterations).

\item Each science image is astrometrically calibrated.  As described by
\citet{lan2023}, this involves fitting parameters of an affine transformation
and a seventh-order geometric distortion polynomial in the TPV projection to
pixel coordinates of sources detected in the image and celestial coordinates of
sources contained in the Gaia DR3 catalog \citep{gai2017, gai2018, gai2021,
gai2022}.  The astrometric calibrations exhibit systematic differences between
the transformed pixel and celestial coordinates of $\lesssim 0.1$ arcsec.

\item Each science image is processed using MaxiMask \citep{pai2020}, which is
a convolutional neural network that identifies contaminants in astronomical
images, including cosmic-ray events and satellite trails.  Pixels flagged by
MaxiMask are excluded from the subsequent analysis.  Each science image is also
processed using MaxiTrack \citep{pai2020}, which is a convolutional neural
network that identifies images affected by tracking errors.

\item For each science image, an additional pixel mask is constructed,
identifying pixels that are found in the master bias image to exhibit
significant effects of random telegraph noise \citep[e.g.][]{cha2019}.  Pixels
flagged in this way are excluded from the subsequent analysis.

\item Each science image is photometrically calibrated.  As described by
\citet{lan2023}, this involves comparing aperture photometry of sources
detected in the image to Sloan $g'$ magnitudes of sources contained in the Gaia
DR3 catalog \citep{gai2017, gai2018, gai2021, gai2022}.  The resulting
magnitude zero points are used subsequently to assess the quality of the
science images.  Note that this procedure scales the luminance images to Sloan
$g'$ magnitudes, although the luminance bandpass is actually roughly comparable
to the sum of the Sloan $g'$ and $r'$ bandpasses.  This introduces a
color-dependent ambiguity in the photometric calibration, which for
low-redshift galaxies amounts to $\approx 0.25$ mag.

\item Each science image is associated with an uncertainty image, which
propagates the $1 \sigma$ uncertainty appropriate for each pixel, starting from
read noise and photon noise.

\item Science images are rejected from the analysis based on (1) poor or
impossible astrometric calibration (indicating clouds or obstruction by an
observatory wall), (2) large width of the autocorrelation function (indicating
out-of-focus images or poor seeing conditions), (3) high background (indicating
substantial man-made or Moon light), (4) low sky transparency (indicating fog,
haze, or clouds), or (5) significant tracking errors (indicating substantial
wind buffeting).

\item The science images are then drizzled \citep{gon2012} onto a common
coordinate grid and coadded weighted for maximum sensitivity in the
background-limited regime according to the uncertainty images.

\end{enumerate}

The resulting coadded images are show in Figures 1 through 6, and a coadded
mosaic of the six images of the entirety of the NGC 5866 Group is shown in
Figure 7.  The measured point-source FWHM and point-source ($5 \sigma$) and
surface-brightness ($3 \sigma$ over $10 \times 10$ arcsec$^2$ regions)
sensitivities of the various images (determined near the centers of the images)
are presented in Table 2.  Note that the FWHM of Table 2 include the combined
effects of focus, seeing, tracking errors, and astrometric errors averaged over
many images.  Also note that the surface-brightness sensitivities of Table 2
are formal statistical values determined from the uncertainty images and
neglect systematic uncertainties associated with field flattening, background
subtraction, scattered starlight, and undetected faint sources.  And finally
note that the sensitivities of Table 2 do not scale in a simple way with
exposure time.  For a telescope like Condor that obtains observations spanning
long dwell times, there will of course be significant variations in seeing,
background, and sky transparency over the course of the (perhaps substantial)
duration of the observations.  So for this reason, exposure time alone is not a
good indicator of the depth of an image.

\begin{figure*}
\centering
\subfloat{
  \includegraphics[width=0.78\linewidth, angle=0]{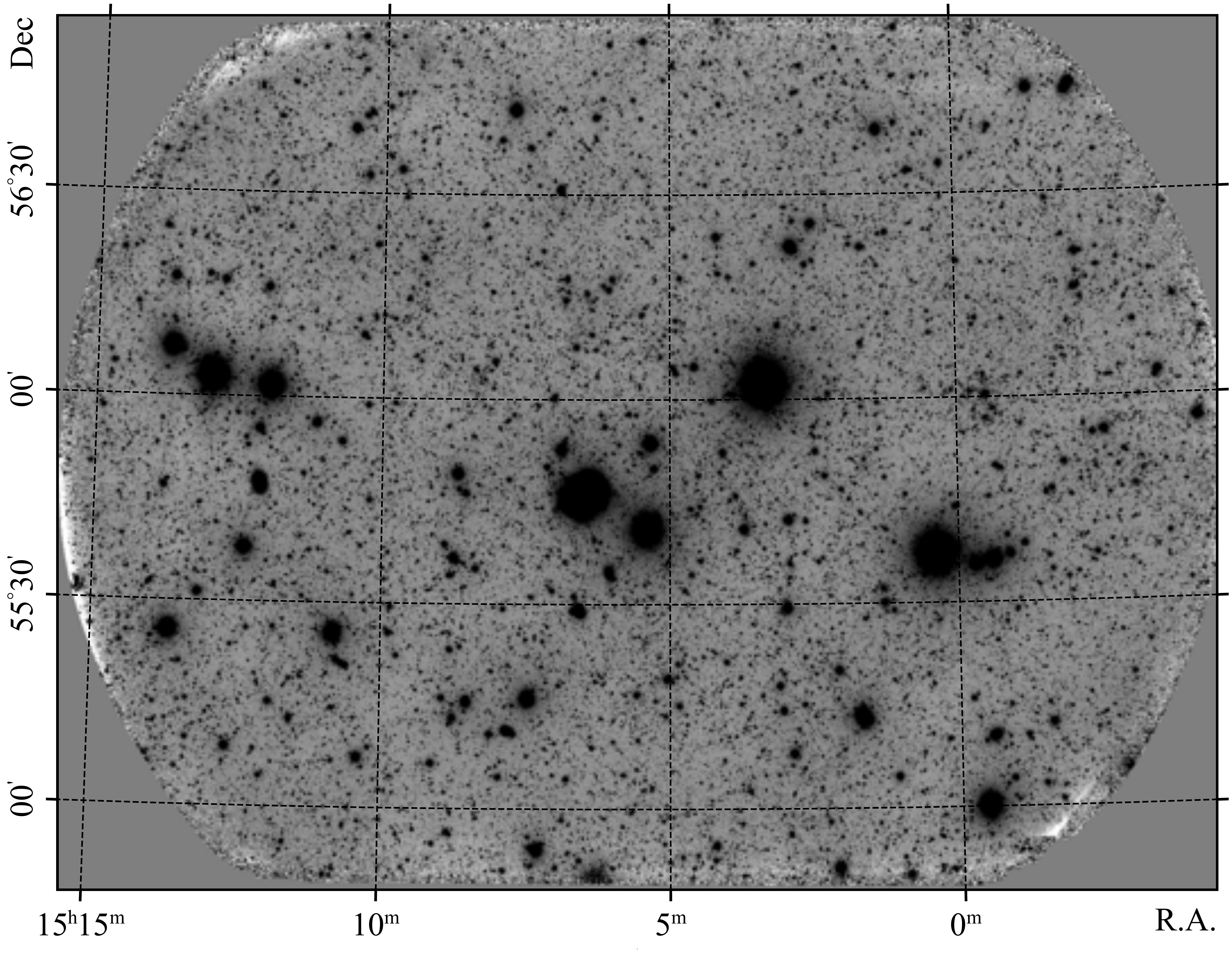}
}
\caption{Coadded image of Condor field 6089.  Image is smoothed by Gaussian
kernel of ${\rm FWHM} = 2.5$ pix.  Total exposure time is 22.7 hr.}
\end{figure*}

\begin{figure*}
\centering
\subfloat{
  \includegraphics[width=0.78\linewidth, angle=0]{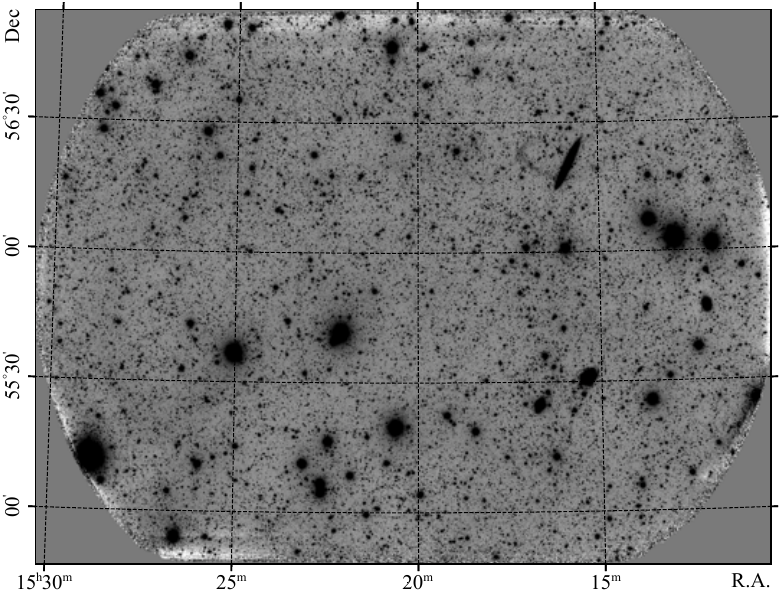}
}
\caption{Coadded image of Condor field 6090.  Image is smoothed by Gaussian
kernel of ${\rm FWHM} = 2.5$ pix.  Total exposure time is 25.0 hr.  Galaxy NGC
5907 is toward upper right of image.}
\end{figure*}

\begin{figure*}
\centering
\subfloat{
  \includegraphics[width=0.78\linewidth, angle=0]{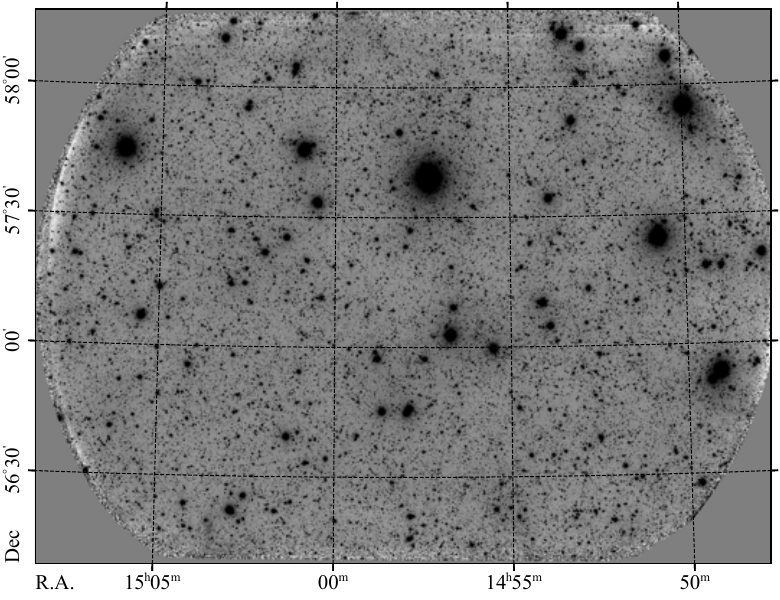}
}
\caption{Coadded image of Condor field 6183.  Image is smoothed by Gaussian
kernel of ${\rm FWHM} = 2.5$ pix.  Total exposure time array is 19.7 hr.}
\end{figure*}

\begin{figure*}
\centering
\subfloat{
  \includegraphics[width=0.78\linewidth, angle=0]{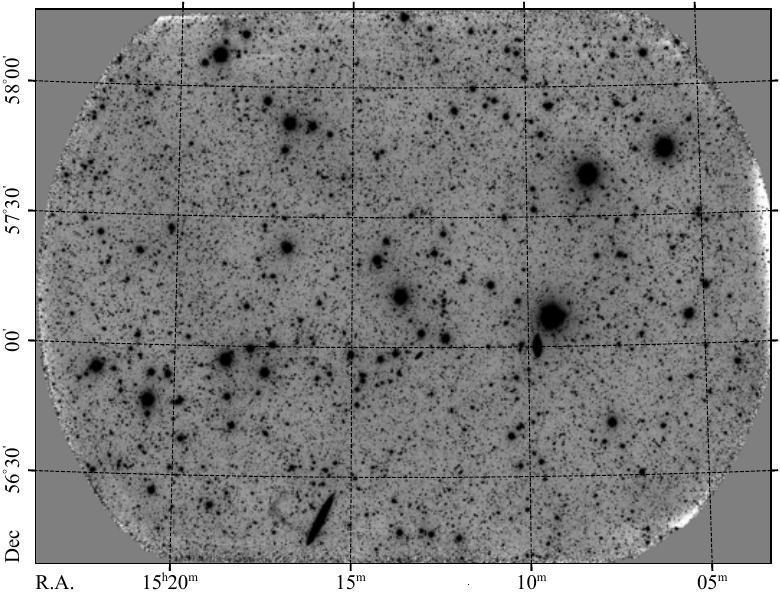}
}
\caption{Coadded image of Condor field 6184.  Image is smoothed by Gaussian
kernel of ${\rm FWHM} = 2.5$ pix.  Total exposure time is 24.1 hr.  Galaxy NGC
5907 is toward lower center of image.}
\end{figure*}

\begin{figure*}
\centering
\subfloat{
  \includegraphics[width=0.78\linewidth, angle=0]{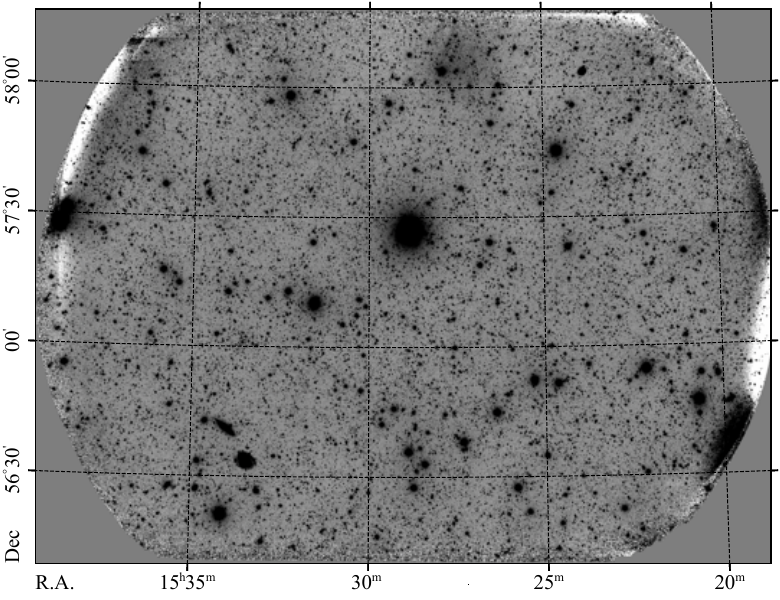}
}
\caption{Coadded image of Condor field 6185.  Image is smoothed by Gaussian
kernel of ${\rm FWHM} = 2.5$ pix.  Total exposure time is 9.1 hr.}
\end{figure*}

\begin{figure*}
\centering
\subfloat{
  \includegraphics[width=0.78\linewidth, angle=0]{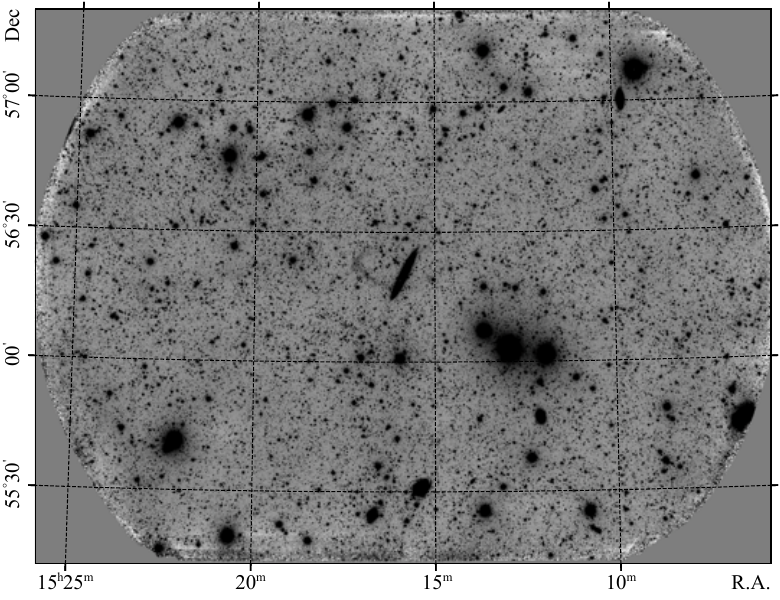}
}
\caption{Coadded image of NGC 5907.  Image is smoothed by Gaussian kernel of
${\rm FWHM} = 2.5$ pix.  Total exposure time is 21.4 hr.  Galaxy NGC 5907 is
near center of image.}
\end{figure*}

\begin{figure*}
\centering
\subfloat{
  \includegraphics[width=1.00\linewidth, angle=-0]{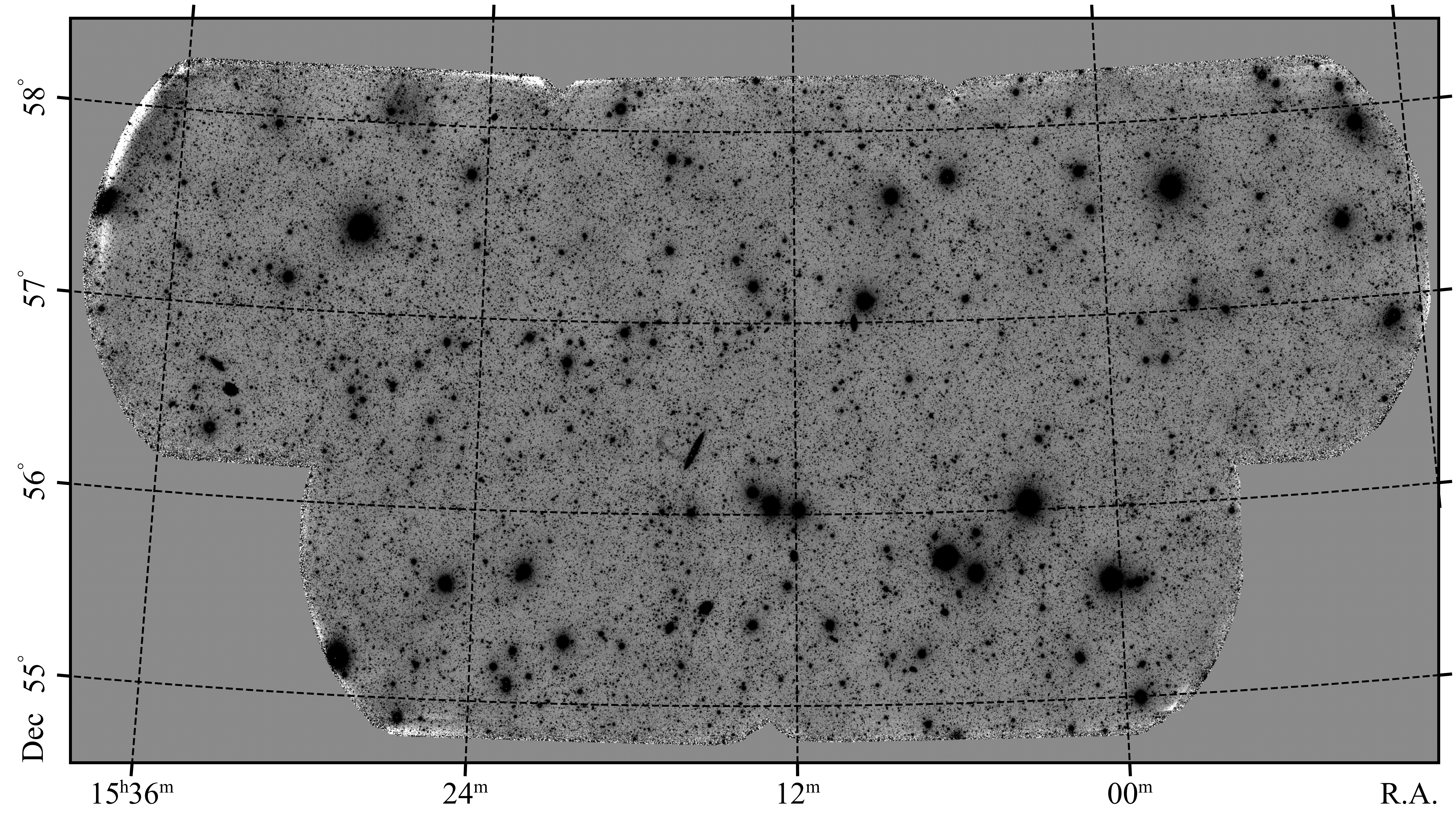}
}
\caption{Coadded mosaic of the six images of the entirety of the NGC 5866
Group.  Image is smoothed by Gaussian kernel of ${\rm FWHM} = 2.5$ pix, and
angular extent of image is $\approx 7.0 \times 3.5$ deg$^2$.  Total exposure
time summed over all six pointings is 122 hr.}
\end{figure*}

\begin{table}
\centering
\caption{Image FWHM and sensitivities.  Point-source sensitivities are $5
\sigma$, and surface-brightness sensitivities are $3 \sigma$ over $10 \times
10$ arcsec$^2$ regions.}
\begin{tabular}{p{1.05in}ccc}
\hline
\multicolumn{2}{c}{ } & \multicolumn{1}{c}{Point} &
\multicolumn{1}{c}{Surface} \\
\multicolumn{1}{c}{ } & \multicolumn{1}{c}{FWHM} &
\multicolumn{1}{c}{Source} & \multicolumn{1}{c}{Brightness} \\
\multicolumn{1}{c}{Pointing} & \multicolumn{1}{c}{(arcsec)} &
\multicolumn{1}{c}{(mag)} & \multicolumn{1}{c}{(mag arcsec$^{-2}$)} \\
\hline
Condor field 6089 \dotfill & 2.6 & 24.9 & 29.5 \\
Condor field 6090 \dotfill & 3.0 & 24.6 & 29.4 \\ % NW
Condor field 6183 \dotfill & 2.6 & 24.8 & 29.4 \\
Condor field 6184 \dotfill & 3.0 & 24.7 & 29.5 \\ % S
Condor field 6185 \dotfill & 3.0 & 23.9 & 28.7 \\
NGC 5907          \dotfill & 2.1 & 25.2 & 29.6 \\ % center
mosaic            \dotfill & 2.3 & 25.5 & 29.9 \\ % center
\hline
\end{tabular}
\end{table}

\section{Assessment of Field Flattening and Background Subtraction}

Errors in field flattening and background subtraction can be significant
sources of systematic uncertainties at low surface-brightness thresholds.  Here
we assess the field flattening and background subtraction of the images of
Figure 1 through 7, concentrating on the mosaic image of Figure 7.

One possible assessment of errors in field flattening and background
subtraction might be obtained by measuring fluctuations within randomly chosen
apertures that by chance are devoid of detectable sources.  But at the faint
limits of the images of Figures 1 through 7, the sky is covered with faint
sources (mostly background galaxies), at an incidence that exceeds 10
arcmin$^{-2}$.  Hence there are essentially {\em no} apertures as large as,
say, $1 \times 1$ arcmin$^2$ (or even $0.5 \times 0.5$ arcmin$^2$) that are
devoid of detectable sources.

Instead, we assess errors in field flattening and background subtraction by
measuring the data covariance over ``background'' pixels of the images, i.e.\
pixels of the images that are {\em not} masked surrounding detectable sources
using NoiseChisel (as described in \S\ 3, enumerated point iii).  On small
spatial scales (i.e.\ on scales of a few pixels), we expect the images to be
highly correlated due to the drizzling process used to coadd the images (as
described in \S\ 3, enumerated point x).  But on larger spatial scales (i.e.\
on scales of tens, hundreds, or thousands of pixels), the images should ideally
exhibit zero covariance, and any non-zero covariance must indicate
large-spatial-scale undulations of the background, which could be due in part
to errors in field flattening and background subtraction.

We consider some region of some image for which the uncertainty image over the
background (i.e.\ unmasked) pixels is roughly constant.  (This applies over the
central regions of all of the images considered here.)  The values of these
pixels can be considered a random variable of zero mean and constant variance.
We write the data covariance $C_l^2$ at some pixel lag $l$ as
\begin{equation}
C_l^2 = \frac{1}{N-1} \sum_i x_i x_{i-l},
\end{equation}
where the sum extends over the $N$ background pixels of the region.  The data
covariance $C_0^2$ at zero pixel lag
\begin{equation}
C_0^2 = \frac{1}{N-1} \sum_i x_i^2
\end{equation}
is the pixel-to-pixel variance of the region.  We further write the correlation
coefficient $\rho_l$ at pixel lag $l$ as
\begin{equation}
\rho_l = \frac{C_l^2}{C_0^2}.
\end{equation}
Here we consider results obtained from a $5000 \times 5000$ pix$^2$ region of
the mosaic image of Figure 7 centered on NGC 5907, although similar results can
of course be obtained using other regions of other images.

The distribution $\Phi(f_\nu)$ of pixel-to-pixel energy fluxes $f_\nu$ of the
background pixels of the mosaic region is shown by the blue curve in Figure 8.
The pixel-to-pixel variance of the mosaic region is measured to be
\begin{equation}
C_0^2 = 1.569 \times 10^{-4} \ \mu{\rm Jy}^2,
\end{equation}
while the median ``statistical'' variance $\sigma_s^2$ of the mosaic region
determined from the background pixels of the uncertainty is image is measured
to be
\begin{equation}
\sigma_s^2 = 2.329 \times 10^{-4} \mu{\rm Jy}^2.
\end{equation}
The background pixels of the uncertainty image are indeed roughly constant over
the mosaic region, and we use the median only to mitigate possible effects of
deviant pixels.  As expected, the pixel-to-pixel variance is {\em less} than
the median variance determined from the uncertainty image, because the
drizzling process used to coadd the images combines nearby pixels, which has
the effect of ``smoothing'' the image and thus reducing the variance.  We
characterize the relationship between the pixel-to-pixel variance and the
median variance determined from the uncertainty image by the ratio
\begin{equation}
r = \frac{C_0^2}{\sigma_s^2} = 0.674.
\end{equation}
Gaussian distribution functions of standard deviations $\sigma_s$ and
$(C_0^2)^{1/2}$ are shown by the orange and green curves, respectively, in
Figure 8.  It is clear that a standard deviation $\sigma_s$ is too wide to
adequately describe the observed distribution (for the reasons described above)
and that a standard deviation $(C_0^2)^{1/2}$ provides a better but still
inadequate description of the observed distribution.  Specifically, the
observed distribution deviates from a Gaussian distribution function due to an
extended tail of positive energy fluxes, which we attribute to sources missed
by the masking procedure.  A standard deviation $0.00995$ $\mu$Jy (determined
by measuring the standard deviation of the observed distribution truncated at
0.025 $\mu$Jy) is shown by the purple curve in Figure 8; this distribution
function adequately describes the observed distribution except for the extended
tail of positive energy fluxes.  We conclude that pixel-to-pixel fluctuations
of the mosaic region are well described by a combination of a Gaussian
distribution function of standard deviation $\approx 0.01$ $\mu$Jy and an
extended tail of positive energy fluxes due to sources missed by the masking
procedure, which is prominent beyond $\approx 2.5$ standard deviations.

\begin{figure}
\centering
\subfloat{
  \includegraphics[width=1.00\linewidth]{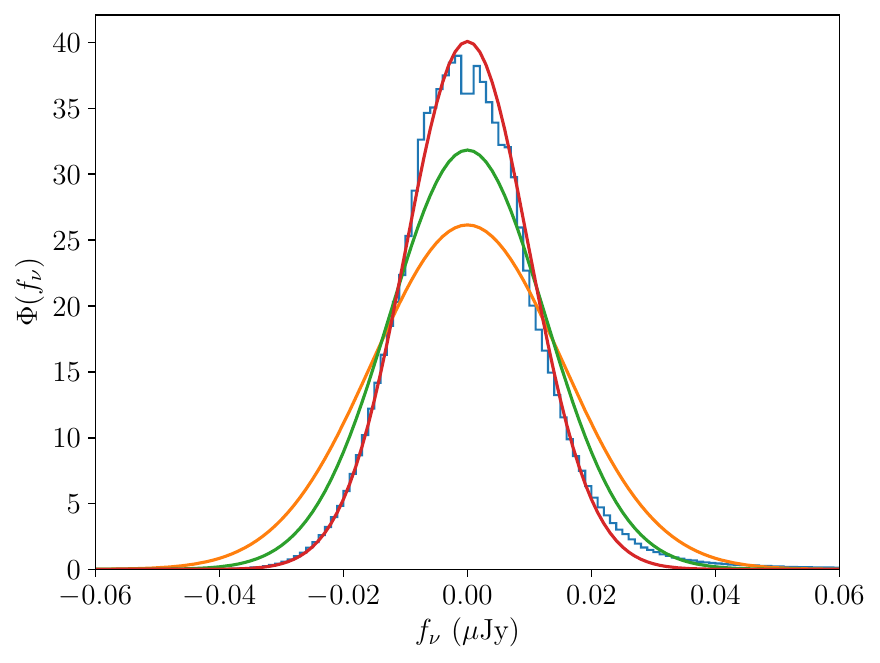}
}
\caption{Distributions $\Phi(f_\nu)$ of pixel-to-pixel energy fluxes $f_\nu$ of
background pixels of mosaic region.  Blue curve shows observed distribution,
and orange, green, and purple curves show Gaussian distribution functions of
standard deviations $\sigma_s$, $(C_0^2)^{1/2}$, and $0.00995$ $\mu$Jy,
respectively.}
\end{figure}

The correlation coefficient $\rho_l$ of the mosaic region for pixel lags over
the interval $l = 0$ through $1000$ is shown in Figure 9.  We note several
results from Figure 9 as follows:  (1) Neighboring pixels are highly
correlated, with correlation coefficients ranging from $\rho_1 = 0.60$ for
immediately adjacent neighbors to $\rho_5 = 0.16$ to $\rho_{10} = 0.04$ to
$\rho_{20} = 0.02$.  We attribute the strong correlation of neighboring pixels
to the drizzling process.  (2) Pixels remain correlated to a pixel lag of $l
\approx 300$, with a correlation coefficient over the range $l = 50$ to 300 of
$\rho_l \approx 0.005$.  We attribute the correlation of pixels at pixel lags
$l = 50$ to 300 to large-spatial-scale undulations of the background.  And (3)
pixels at pixel lags $l \gtrsim 300$ are uncorrelated or only weakly
correlated.

\begin{figure}
\centering
\subfloat{
  \includegraphics[width=1.00\linewidth]{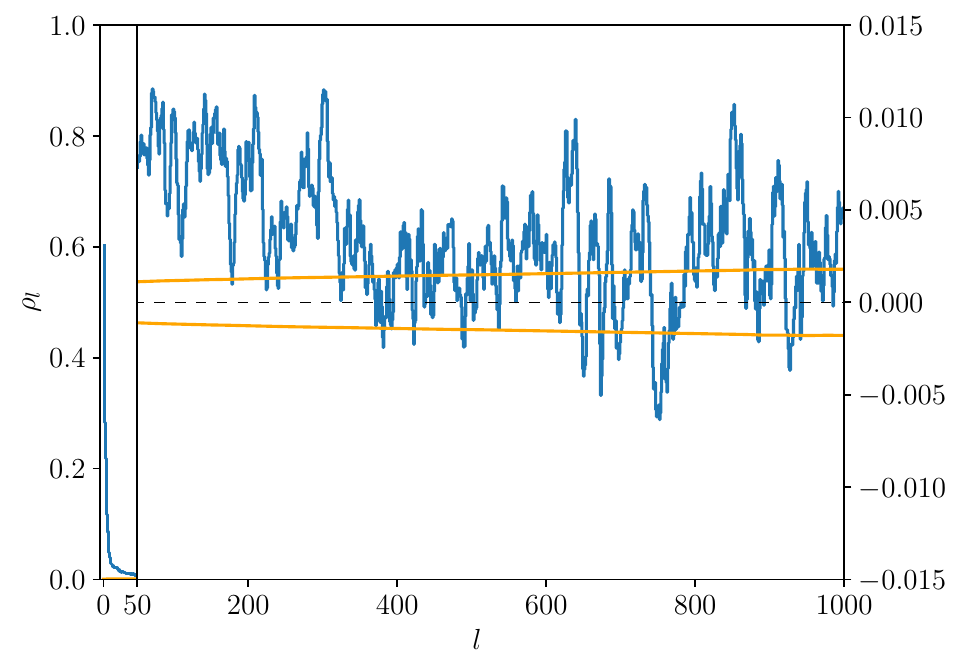}
}
\caption{Correlation coefficient $\rho_l$ (blue curves) together with positive
and negative one standard deviation uncertainties (orange curves) of mosaic
region for pixel lags over intervals $l = 0$ through $50$ (left panel and left
scale) and $l = 50$ to $1000$ (right panel and right scale).}
\end{figure}

We now consider the fluctuations attributable only to background within an
aperture that encompasses $N$ pixels.  If the $N$ pixels are uncorrelated, then
the variance $\sigma_N^2$ of the background within the aperture is
\begin{equation}
\sigma_N^2 = \sum_i C_0^2 = N C_0^2,
\end{equation}
where the sum extends over the pixels that comprise the aperture.  If the $N$
pixels are correlated, then the variance is
\begin{equation}
\sigma_N^2 = \sum_i C_0^2 + \sum_i \sum_{j \ne i} C_l^2 \approx N C_0^2 + N
C_0^2 2 \pi \sum_{l=1}^{n/2} l \rho_l,
\end{equation}
where the sums over $i$ and $j$ extend over the pixels that comprise the
aperture and the sum over $l$ extends over the diameter $n \sim N^{1/2}$ of the
aperture.  Expressing $C_0^2$ in terms of $\sigma_s^2$ and $r$ via equation (6)
then yields
\begin{equation}
\sigma_N^2 = \left( 1 + 2 \pi \sum_{l=1}^{n/2} l \rho_l \right) N r \sigma_s^2.
\end{equation}
The corresponding relationship expressed in terms of a standard deviation
rather than a variance is
\begin{equation}
\sigma_N = \left( 1 + 2 \pi \sum_{l=1}^{n/2} l \rho_l \right)^{1/2} N^{1/2}
r^{1/2} \sigma_s.
\end{equation}
Equation (10) provides the way to relate the formal statistical uncertainties
of the uncertainty images (and hence the sensitives presented, e.g.\, in Table
2) to the actual uncertainties including effects of pixel-to-pixel correlations
on small spatial scales (due to the drizzling process) and on large spatial
scales (due to undulations of the background).  Specifically, the ultimate
effect of pixel-to-pixel correlations is to alter the standard deviation
attributable only to background of an aperture that encompasses $N$ pixels by a
factor $f_N$ given by
\begin{equation}
f_N = \left( 1 + 2 \pi \sum_{l=1}^{n/2} l \rho_l \right)^{1/2} r^{1/2}
\end{equation}
with respect to the value
\begin{equation}
\sigma_N = N^{1/2} \sigma_s
\end{equation}
that is obtained by considering the uncertainty images alone.

The resulting values of $f_N$ measured for the mosaic region are shown (on a
magnitude scale) versus angular scale $\theta$ (i.e.\ expressing diameter $n$
in angular units) in Figure 10, using the correlation coefficient $\rho_l$ from
Figure 9 and the ratio $r$ from equation (6).  Fluctuations on single-pixel
scales are {\em less} by a factor $0.82$ (or $-0.2$ mag) than the value
obtained by considering the uncertainty image alone due to the drizzling
process.  Fluctuations of the background over $0.5 \times 0.5$ and $1 \times 1$
arcmin$^2$ apertures exceed the values obtained by considering the uncertainty
image alone by around $2.0$ and $2.3$ mag, respectively.

The values shown in Figure 10 represent {\em upper limits} to the errors in
field flattening and background subtraction of the mosaic region, because
fluctuations in the background also arise due to scattered starlight and to
undetected faint sources (including the faint sources that make up the extended
tail of positive energy fluxes of the distribution of pixel-to-pixel energy
fluxes of the background pixels of the mosaic regions shown in Figure 8).
Further, fluctuations in the background are not necessarily simply related to
sensitivity; for example, a source of diameter $0.1$ arcmin might be detected
despite undulations in the background on scales of $1$ arcmin.  A detailed
accounting of all sources of fluctuations in the background will be described
elsewhere.

\begin{figure}
\centering
\subfloat{
  \includegraphics[width=1.00\linewidth]{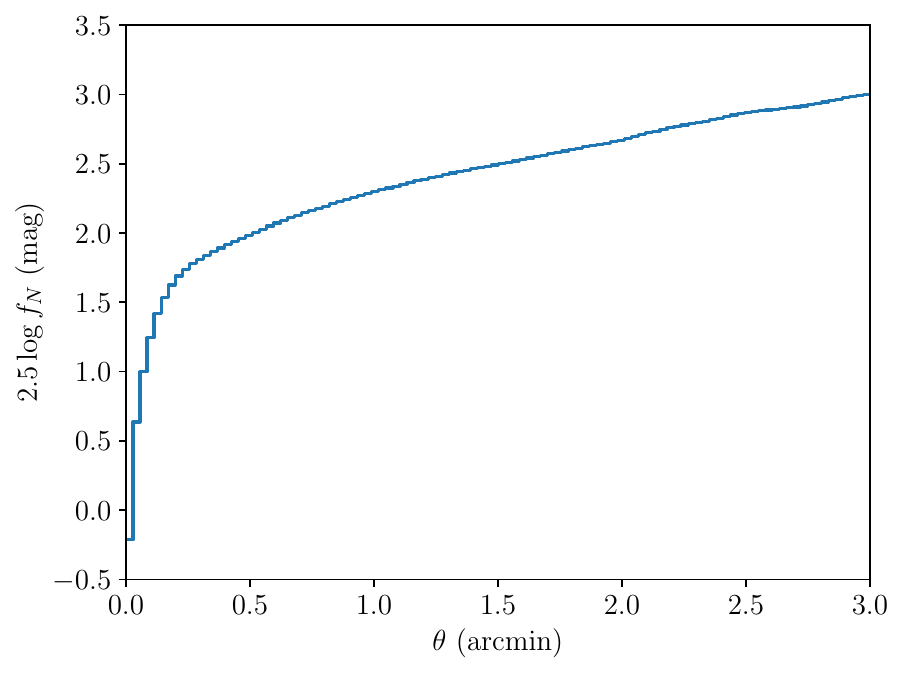}
}
\caption{Factor $f_N$ by which standard deviation of background is altered with
respect to value obtained considering uncertainty images alone (on a magnitude
scale) of the mosaic region versus angular scale $\theta$.}
\end{figure}

\section{Correction for Scattered Starlight}

Although Condor exhibits a very clean point-spread function
\citep[PSF,][]{lan2023}, scattered starlight can be a significant source of
systematic noise.  Hence to fully exploit the sensitivity of the images
described in \S\ 3 to very low-surface-brightness features, it is necessary to
correct for scattered starlight by (1) accurately determining the PSF on large
angular scales and (2) using the resulting PSF to model and subtract the
contributions of all stars within (and possibly even beyond) the field of view.
Details of our method of PSF determination and subtraction will be described
elsewhere, but here we present a brief summary of the procedures and results.

To determine the PSF, we compare a ``data'' image with a ``model'' image, where
we take the model image to be the convolution of a ``sky'' image with a ``PSF''
image.  We allow for the possibility that the model image (and hence the PSF
image) is expressed on a finer grid than the data image (i.e.\ is
``subsampled'' with respect to the data image).  We assume that the sky
consists only of stars (i.e.\ we mask regions around galaxies and other
non-stellar sources), and hence we take the sky image to be a sum of delta
functions, where the locations of the delta functions (i.e.\ the locations of
the stars) are taken as given.  We then write the comparison between the data
image and the model image as a linear least squares problem, and we solve the
normal equations \citep[e.g.][]{pre2007} to minimize $\chi^2$ with respect to
some parameters.  In particular, if the normalizations of the delta functions
(i.e.\ the energy fluxes of the stars) are taken as given, then we solve the
normal equations for the PSF image, or if the PSF image is taken as given, then
we solve the normal equations for the energy fluxes of the stars.  In practice,
starting with any reasonable guess for the energy fluxes of the stars and
iterating between solving for the PSF image and solving for the energy fluxes
of the stars, the solution quickly converges to the desired simultaneous
solution.

Our primary objective is to determine and subtract the PSF on large angular
scales, and for this purpose, the limitations of a pixel-based approach are
obvious:  Near the core of the PSF, a fine pixel grid is both necessary
(because the PSF exhibits rapid variations at small angles) and feasible
(because observations of the PSF contain substantial signal at small angles).
But moving outward from the core of the PSF, the same fine pixel grid becomes
both unnecessary (because the PSF exhibits less rapid variations at larger
angles) and implausible (because observations of the PSF contain less signal at
larger angles).  Clearly some some sort of adaptive parametrization is
required, which is finer near the core of the PSF and grows increasingly
coarser moving outward.

Accordingly, we modify the method described above to allow arbitrary groupings
of pixels on the pixel grid of the model image (and hence the PSF image) to be
treated as single parameters.  Specifically, we rewrite the normal equations to
allow for (1) a pixellated parameter grid near the core of the PSF and (2) a
circular annulus (if azimuthal symmetry is assumed) or annulus sector (if
azimuthal symmetry is not assumed) parameter grid moving outward from the core.
Together these modifications optimally represent the PSF over a huge dynamic
range, vastly reduce the dimensionality of the problem, and remain linear in
the parameters.

We emphasize that the method determines the PSF by {\em simultaneously} fitting
all stars in the field, so there is no requirement of incorporating only
isolated stars into the analysis.

In practice, we take locations and starting values of the energy fluxes of the
stars from the Gaia DR3 catalog \citep{gai2017, gai2018, gai2021, gai2022}, and
we solve for the PSF image and the energy fluxes of the stars assuming a
pixellated PSF at angular radius $\theta < 20$ arcsec and an
azimuthally-symmetric PSF at angular radius $20$ arcsec $< \theta < 10$ arcmin,
masking regions around galaxies and other non-stellar sources.  We then
subtract the model from the data, masking pixels of the result near the very
cores of the stars at an isopohotal flux limit.  (This masking is necessary
because residuals near the cores of the stars can be large compared with the
very low surface brightness limits farther from the cores.)  A radial cut of a
representative example of the PSF determined from the mosaic image is shown in
Figure 11.  It is apparent from Figure 11 that the ``aureole'' portion of the
PSF (i.e.\ beyond an angular radius $\theta \approx 20$ arcsec) roughly follows
a $\theta^{-2}$ radial profile, which is similar to the radial profiles of some
other telescopes used for low-surface-brightness imaging
\citep[e.g.][]{san2014}.  The processed image of a portion of the mosaic image
surrounding NGC 5907 obtained by modeling and subtracting the contributions of
stars in the Gaia DR3 catalog is show in two different stretches in Figure 12;
Figure 12 also shows schematic representation of features described in \S\
6 below.

\begin{figure}
\centering
\subfloat{
  \includegraphics[width=1.00\linewidth]{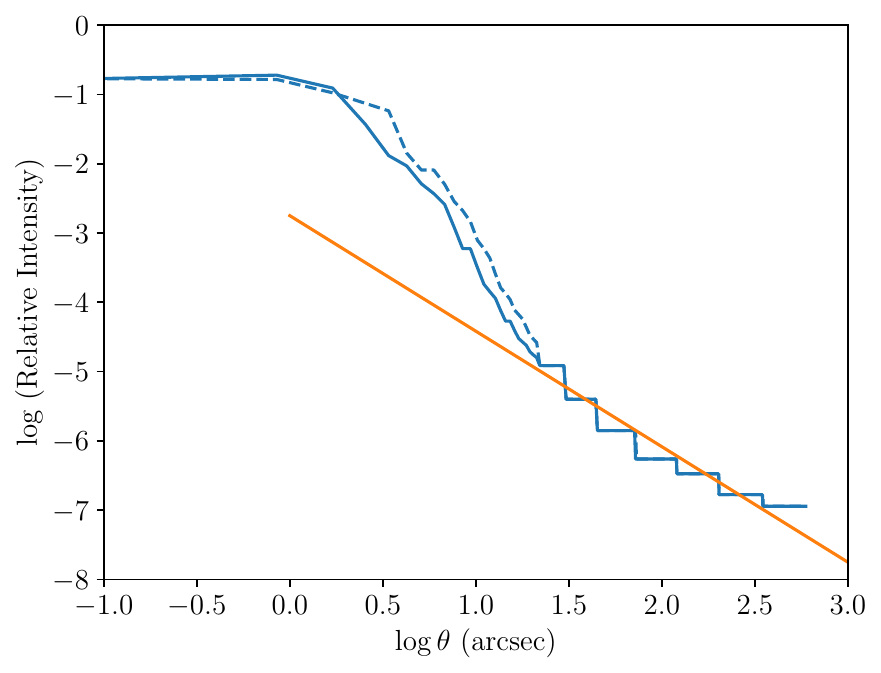}
}
\caption{Radial cut of representative example of PSF determined according to
description of \S\ 4, normalized to unit area.  Solid and dashed curves show
PSF in two opposite directions.  Orange line segment show $\theta^{-2}$ radial
profile.}
\end{figure}

\begin{figure*}
\centering
\subfloat{
  \includegraphics[width=1.00\linewidth]{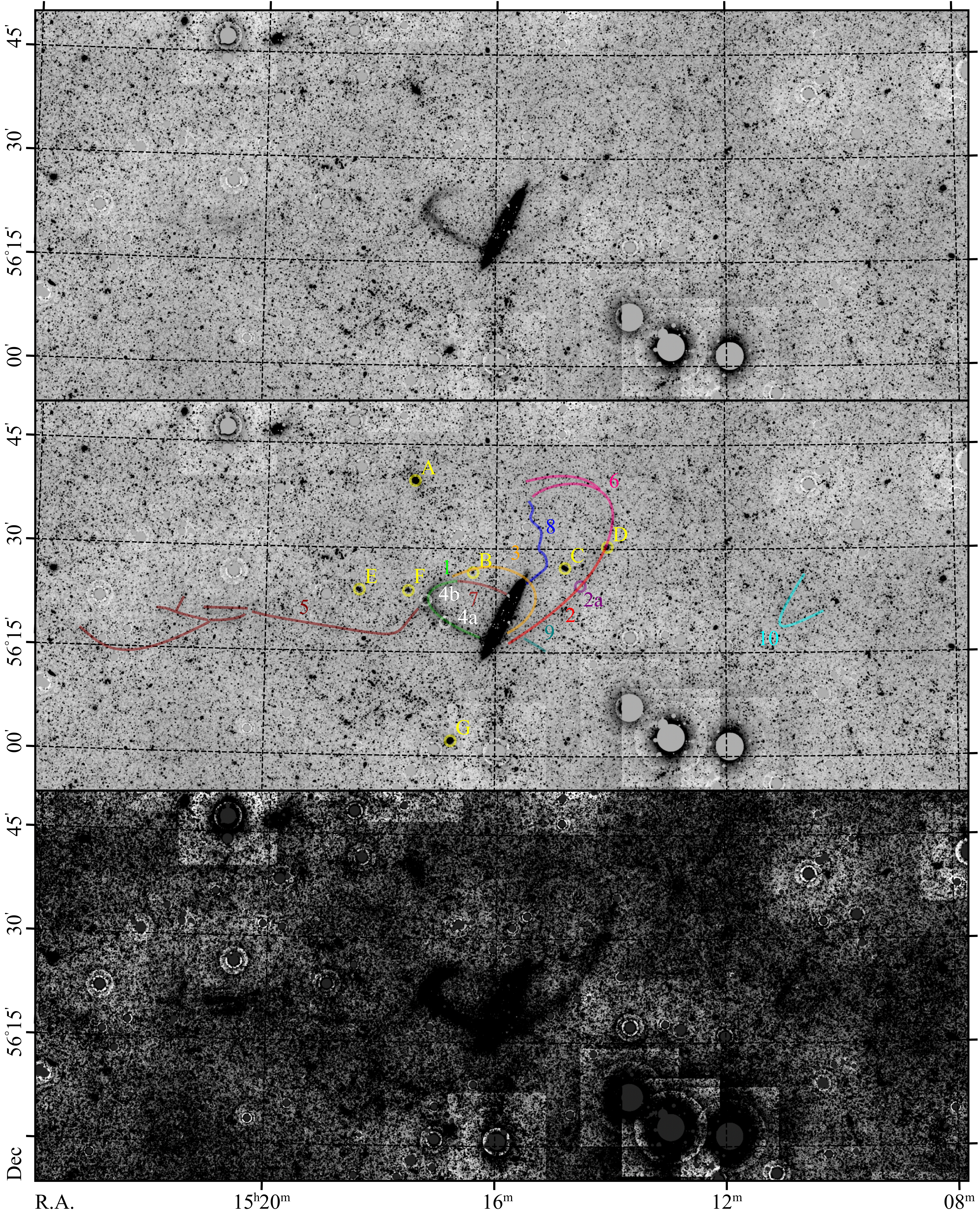}
}
\caption{Processed mosaic image of region around NGC 5907 at shallower (top and
middle panels) and deeper (bottom panel) stretches.  Image is smoothed by
Gaussian kernel of ${\rm FWHM} = 2.5$ pix.  Middle panel shows schematic
representation of features described in \S\ 6 labeled as follows:  1 (green)
eastern stream, 2 (red) western stream, 2a (purple) apparent gap in western
stream, 3 (orange) putative second loop of stellar stream, 4a and 4b (white)
feature and clump of sources in eastern stream, 5 (maroon) linear feature
terminating on patch, 6 (pink) putative extension of western stream, 7 (brown)
continuation of eastern stream, 8 (blue) western ``horn'', 9 (turquoise)
southern ``spur,'' 10 (aqua) western ''hook,'' and A--G (yellow) dwarf
galaxies.}
\end{figure*}

Our analysis differs from the analysis of \citet{van2019} in that we model and
subtract only contributions from stars (and the occasional galaxy) that are
contained in the Gaia DR3 catalog whereas they model and subtract contributions
from all ``compact emission sources.''  It is apparent from Figure 12 that our
processed images exhibit a large number of faint sources, the vast majority of
which are faint, background galaxies.  But some fraction of these faint sources
might be associated with NGC 5907, e.g.\ as dwarf galaxies, globular clusters,
or perhaps other types of star clusters or associations.  This difference
between our analysis and the analysis of \citet{van2019} leads to some
important consequences, as is described below.

\section{Results and Comparison With Previous Work}

Here we use the processed mosaic image of the region surrounding the galaxy NGC
5907 shown in Figure 12 to assess the various established, proposed, and
tentative features reported by others and propose new features and new
interpretations of some previously-reported features.

\subsection{Eastern Stream}

Our image of the eastern stream (which is indicated as feature 1 in green in
Figure 12) through the luminance filter is consistent in location, size, shape,
brightness, and overall morphology with the image of the eastern stream through
the sum of the Sloan $g'$ and $r'$ filters presented by \citet{van2019}, which
we established by overlaying and comparing the two images.  In contrast, both
our image and the image of \citet{van2019} of the eastern stream are
inconsistent with the image of the eastern stream through the sum of the $R$,
$G$, $B$, and luminance filters presented by \citet{mar2010} in the sense that
the portion of the stream that is maximally displaced from the disk of the
galaxy (i.e.\ the apex of the stream) in the image of \citet{mar2010} lies {\em
interior} to the same portion of the stream in our image and the image of
\citet{van2019}, which we established by overlaying and comparing the three
images.  The displacement of the stream over this region between the two
dichotomous sets of images amounts to $\approx 1$ arcmin.  This discrepancy in
the location of the eastern stream was noted previously by \citet{van2019}.

We measured the typical surface brightness through the luminance filter of the
eastern stream to be $\mu_{\rm lum} \approx 27.4$ mag arcsec$^{-2}$.  This
value may be compared with the peak surface brightness through the Sloan $g'$
filter of the eastern stream measured by \citet{van2019} to be $\mu_{g'} =
27.6$ mag arcsec$^{-2}$.

\subsection{Western Stream}

Our image of the western stream (which is indicated as feature 2 in red in
Figure 12) through the luminance filter is consistent in location, size, shape,
brightness, and overall morphology with the image of the western stream through
the Sloan $g'$ and $r'$ filters presented by \citet{van2019}, which we
established by overlaying and comparing the two images.  In contrast, both our
image and the image of \citet{van2019} of the western stream are inconsistent
with the image of the western stream through the $R$, $G$, $B$, and luminance
filters presented by \citet{mar2010} in the sense that the image presented by
\citet{mar2010} shows only a small portion of the western stream, near where it
emerges from the southern edge of the galaxy, and does not show the remainder
of the stream, as it bends toward the north, which we established by overlaying
and comparing the three images.  This discrepancy in the morphology of the
western stream was noted previously by \citet{van2019}.

We measured the typical surface brightness through the luminance filter of the
western stream to be $\mu_{\rm lum} \approx 28.6$ mag arcsec$^{-2}$.  This
value may be compared with the typical surface brightness through the Sloan
$g'$ filter of the western stream measured by \citet{van2019} to be $\mu_{g'} =
28.8$ mag arcsec$^{-2}$.  Thus, consistent with results of \citet{van2019}, we
find that the western stream is of surface brightness significantly lower than
that of the eastern stream (by $\approx 1.2$ mag arcsec$^{-2}$).  Apparently,
images that fail to detect all or part of the western stream must not reach
surface-brightness sensitivities of $\approx 28.7$ mag arcsec$^{-2}$ over
angular scales necessary to detect the stream.

Our image also shows an apparent gap in the western stream due east of the
galaxy.  The gap is followed by a marked thickening or enhancement of the
western stream, although its surface brightness does not increase significantly
in this thicker region.  The apparent gap in the western stream is indicated as
feature 2a in purple in Figure 12.  The gap extends $\approx 70$ arcsec, which
at the distance of NGC 5907 corresponds to $\approx 6$ kpc.

\subsection{Putative Second Loop of Stellar Stream}

Neither our image through the luminance filter nor the image through the Sloan
$g'$ and $r'$ filters presented by \citet{van2019} show any evidence at all of
the second loop of the stellar stream seen in the image through the $R$, $G$,
$B$, and luminance filters presented by \citet{mar2010}.  The location of the
putative second loop of the stellar stream is indicated as feature 3 in orange
in Figure 12, which we determined by overlaying and tracing the feature from
the image of \citet{mar2010}.  Our image reaches a formal $3 \sigma$
surface-brightness sensitivities over $10 \times 10$ arcsec$^2$ regions of
$\approx 29.9$ mag arcsec$^{-2}$ (see Table 2), and \citet{van2019} quote a $3
\sigma$ surface-brightness sensitivity of 29.4 mag arcsec$^{-2}$ (although they
do not state the angular scale over which this limit is meant to apply).  We
see no plausible way for color effects to explain the discrepancy, given that
the observations reported by \citet{mar2010} were obtained either through the
luminance filter, as were our observations, or through a ``synthetic''
luminance filter (formed using observations obtained through the $R$, $G$, and
$B$ filters).

We conclude that the second loop of the stellar stream seen in the image
presented by \citet{mar2010} is not real and must result from some artifact of
their data processing; we further suggest that the discrepancies in the
location of the eastern stream and the morphology of the western stream must
also result from some artifact of their data processing.

\subsection{Putative Remnant of Nearly Disrupted Progenitor Galaxy and
Luminosity-Weighted Midpoint of Eastern Stream}

Our image shows a ``feature'' (which is indicated as feature 4a in white in
Figure 12) near the location of the ``density enhancement near the
luminosity-weighted midpoint of the [eastern] stream'' noted by
\citet{van2019}.  But our images resolve this feature into a clump of sources,
which we interpret as members of a background galaxy group or cluster rather
than as the ``likely remnant of a nearly disrupted progenitor galaxy'' proposed
by \citet{van2019}.

Our image also shows a different clump of sources (which is indicated as
feature 4b in white in Figure 12) near the ``luminosity-weighted midpoint'' of
the eastern stream, including one relatively bright galaxy that might be a
dwarf galaxy associated with NGC 5907 or might be a member of a background
galaxy group or cluster.  (This galaxy is included into the Gaia DR3 catalog,
and our analysis described in \S\ 4 attempted to model and subtract it,
although unsuccessfully since it is not a point source.)  But in either case,
it is clear from Figure 12 that galaxies (background or otherwise) or other
discrete sources contribute significantly to the luminosity-weighted midpoint
of the eastern stream.  In particular, much of the ``density enhancement'' of
the eastern stream found by \citet{van2019} (i.e.\ the portions of their Figure
3 depicted in red) is in fact contributed by discrete sources, which we
established by overlaying and comparing our Figure 12 with their Figure 3.
(The discrete sources can be picked out one by one by means of this
comparison.)  Further, the possible asymmetry in the density enhancement of the
eastern stream noted by \citet{van2019} is in fact contributed by discrete
sources, including in particular the relatively bright galaxy noted above.

We conclude that the feature proposed by \citet{van2019} as the likely remnant
of a nearly disrupted progenitor galaxy is not the progenitor galaxy but is in
fact a member of a background galaxy group or cluster and that the density
enhancement and possible asymmetry of the density enhancement of the eastern
stream noted by \citet{van2019} is in fact contributed by discrete sources.
This difference of interpretation presumably arises due to the higher angular
resolution of our observations in comparison with the observations of
\citet{van2019}.

\subsection{Linear Feature Terminating on Patch}

Our images confirm the ``linear'' feature emanating from the eastern stream
toward the east and terminating on a ``patch'' of emission identified by
\citet{van2019}.  But our images further indicate that the feature continues
past the patch toward the east and eventually terminates on another patch of
emission located $\approx 0.37$ deg away from the first patch (which itself is
located $\approx 0.67$ deg from the center of NGC 5907).  Our images also
further indicate that first patch is itself resolved into two roughly parallel
linear segments running roughly east-west and another clump of emission toward
the east.  This entire structure is indicated as feature 5 in maroon in Figure
12.  In total, the structure stretches $\approx 0.85$ deg from where it
emanates near the apex of the eastern stream to where it terminates on the
second patch.

We measured the typical surface brightness through the luminance filter of the
first patch to be $\mu_{\rm lum} \approx 28.1$ mag arcsec$^{-2}$ and the
typical surface brightness through the luminance filter of the second patch to
be $\mu_{\rm lum} \approx 28.9$ mag arcsec$^{-2}$.  Thus we find that both
patches are of surface brightness significantly lower than that of the eastern
stream and that the first patch is of surface brightness significantly higher
than that of the western stream while the second patch is of surface brightness
comparable to that of the western stream.  The linear feature and the
continuation of the linear feature vary significantly in brightness along their
lengths, but we measured a typical surface brightness through the luminance
filter of these features to be $\mu_{\rm lum} \approx 29.7$ mag arcsec$^{-2}$.
Thus we find that the linear feature and the continuation of the linear feature
are typically of surface brightness significantly lower than that of the
patches (by $\approx 1.0$ mag arcsec$^{-2}$), although we note a significant
brightening of the linear feature west of the first patch, roughly midway
between the first patch and the eastern stream.  We measured the angular extent
of the first patch to be $\approx 530 \times 240$ arcsec$^2$ and the angular
extent of the second patch to be $\approx 220 \times 270$ arcsec$^2$, where the
measurements apply to an isophotal contour of $\approx 29$ mag arcsec$^{-2}$.

We conclude that the linear feature emanating from the eastern stream toward
the east and terminating on a patch identified by \citet{van2019} are part of
a yet larger structure.  If this structure is at the distance of NGC 5907
(which is plausible or likely given that it appears to emanate near the apex of
the eastern stream), then the first patch is located $\approx 200$ kpc from the
center of the galaxy, the second patch is located $\approx 300$ kpc from the
center of the galaxy, and the entire structure stretches $\approx 240$ kpc from
where it emanates near the apex of the eastern stream to where it terminates on
the second patch.  Further, the spatial extent of the first patch is $\approx
43 \times 20$ kpc$^2$, the spatial extent of the second patch is $\approx 18
\times 22$ kpc$^2$, the absolute magnitude through the luminance filter of the
first patch is $\approx -15.4$, i.e.\ roughly $0.6\%$ that of the Milky Way,
and the absolute magnitude through the luminance filter of the second patch is
$\approx -14.2$, i.e.\ roughly $0.2\%$ that of the Milky Way (where we take the
Sloan $g'$ absolute magnitude of the Milky Way to be $-21.0$, e.g.\
\citealp{bla2016}).  Multi-band imaging of the field surrounding NGC 5907 will
be necessary to establish the nature of the patches of emission.

\subsection{Putative Extension of Western Stream}

Our image confirms the extension of the western stream (which is indicated as
feature 6 in pink in Figure 12) tentatively identified by \citet{van2019}.
This extension continues along the direction of the western stream described in
\S\ 6.2 toward the north and then curls back south toward NGC 5907, about 0.3
deg north of the center of the galaxy.  Our image further shows that the stream
appears to bifurcate near its apex.  We measured the typical surface brightness
through the luminance filter of the extension of the western stream to be
$\mu_{\rm lum} \approx 28.9$ mag arcsec$^{-2}$.  Thus we find that the
extension of the western stream is of surface brightness lower than that of the
rest of the western stream (by $\approx 0.3$ mag arcsec$^{-2}$).

\subsection{Putative Continuation of Eastern Stream}

Our image confirms the continuation of the eastern stream (which is indicated
as feature 7 in brown in Figure 12) tentatively identified by \citet{van2019}.
We measured the typical surface brightness through the luminance filter of the
continuation of the eastern stream to be $\mu_{\rm lum} \approx 29.0$ mag
arcsec$^{-2}$.  Thus we find that the continuation of the eastern stream is of
surface brightness significantly lower than that of the bulk of the eastern
stream and lower even than that of the western stream.  There is some
indication of a gap between the brighter bulk of the eastern stream and the
fainter continuation of the eastern stream that joins up to the disk, although
this gap is roughly coincident with three Gaia sources, which muddy the
interpretation.

\subsection{Western ``Horn''}

Our image reveals a new western ``horn'' (which is indicated as feature 8 in
blue in Figure 12) emanating from the western side of the northern portion of
the disk of NGC 5907 and extending to the northwest.  The horn constitutes a
thin, roughly linear feature of diffuse emission.  We also tentatively identify
a continuation of the horn that meanders from the northern tip of the horn
northward by $\approx 0.15$ deg to the extension of the western stream
described in \S\ 6.6.  We measured the typical surface brightness through the
luminance filter $\mu_{\rm lum}$ of the western horn to be $\mu_{\rm lum}
\approx 29.0$ mag arcsec$^{-2}$.  Thus we find that the western horn is of
surface brightness lower than that of the western stream.

The western horn is apparent in the image through the sum of the Sloan $g'$ and
$r'$ filters presented by \citet{van2019}, although these authors did not call
attention to the feature.

\subsection{Southern ``Spur''}

Our image reveals a new southern ``spur'' emanating from the southern portion
of the western stream and continuing to the southwest.  The spur comprises a
band of diffuse emission of thickness comparable to the thickness of the
western stream that runs almost perpendicular to the western stream.  We
measured the typical surface brightness through the luminance filter $\mu_{\rm
lum}$ of the southern spur to be $\mu_{\rm lum} \approx 29.0$ mag
arcsec$^{-2}$.  Thus we find that the southern spur is of surface brightness
lower than that of the western stream.

The southern spur is not obviously evident in the image through the sum of the
Sloan $g'$ and $r'$ filters presented by \citet{van2019}.

\subsection{Western ``Hook''}

Our image reveals a new western ``hook'' (which is indicated as feature 10 in
aqua in Figure 12) located $\approx 0.68$ deg due west of the center of NGC
5907.  Hence the western hook is about as far west of the galaxy as the first
patch described in \S\ 6.5 is east of the galaxy.  There is no clear and
obvious connection between the hook and NGC 5907, but if the hook is at the
distance of the galaxy, then it is located $\approx 200$ kpc from the center of
the galaxy.  We measured the typical surface brightness through the luminance
filter $\mu_{\rm lum}$ of the western hook to be $\mu_{\rm lum} \approx 29.0$
mag arcsec$^{-2}$.  Thus we find that the western hook is of surface brightness
lower than that of the western stream.

The western hook is not covered by the image through the sum of the Sloan $g'$
and $r'$ filters presented by \citet{van2019}.

\subsection{Dwarf Galaxies}

Figure 12 calls attention to several sources, which are identified as sources A
through G in yellow in the figure.  Source B in Figure 12 is the
previously-uncataloged putative dwarf galaxy located just west of the eastern
stream reported by \citet{van2019}.  The proximity of this galaxy to the
eastern stream obviously suggests that the galaxy is associated with (rather
than behind) NGC 5907, but without spectroscopy or multi-band imaging, it is
not possible to know for sure.

\begin{table*}
\centering
\caption{Properties of known galaxies in the immediate vicinity of NGC 5907.}
\begin{tabular}{p{0.5in}ccccccccc}
\hline
\multicolumn{2}{c}{ } &
\multicolumn{2}{c}{J2000} &
\multicolumn{1}{c}{$v_{\rm rec}$} &
\multicolumn{2}{c}{ } &
\multicolumn{1}{c}{Absolute} &
\multicolumn{1}{c}{$b$} &
\multicolumn{1}{c}{$b_{\rm disk}$} \\
\cline{3-4}
\multicolumn{1}{c}{Source} &
\multicolumn{1}{c}{Name} &
\multicolumn{1}{c}{R.A.} &
\multicolumn{1}{c}{Dec} &
\multicolumn{1}{c}{(km s$^{-1}$)} &
\multicolumn{1}{c}{Type} &
\multicolumn{1}{c}{Magnitude} &
\multicolumn{1}{c}{Magnitude} &
\multicolumn{1}{c}{(kpc)} &
\multicolumn{1}{c}{(kpc)} \\
\hline
A \dotfill & MCG+10-22-010 &
15:17:25.2 &
$+$56:39:48.5 &
$781 \pm 80$ & dIrr & $14.987 \pm 0.003$ & $-16.2$ & 117 & 99 \\
C \dotfill & LEDA 54419 &
15:14:47.8 &
$+$56:27:14.8 &
710 & dIrr & $16.2006\pm 0.004$ & $-14.9$ & 56 & 21 \\
D \dotfill & 2MASX J15140431+5630186 &
15:14:04.3 &
$+$56:30:18.7 &
... & ...  & $19.6156\pm 0.008$ & $-11.5$ (?) & 89 & 43 \\
E \dotfill & LEDA 2535522 &
15:18:23.6 &
$+$56:23:58.1 &
... & ...  & $20.44\pm 0.01$ & $-10.7$ (?) & 106 & 102 \\
G \dotfill & LEDA 2523331 &
15:16:46.5 &
$+$56:02:16.0 &
... & ...  & 20.58 & $-10.6$ (?) & 86 & 0 \\
\hline
\end{tabular}
\end{table*}

But interestingly, there are several {\em known} galaxies in the immediate
vicinity of NGC 5907---two of which are {\em known} to be associated with the
galaxy---that were not considered by the analysis of \citet{van2019} because
they were modeled and subtracted by their analysis as ``compact emission
sources.'' These include sources A, C, D, E, and G in Figure 12.  In
particular:
\begin{itemize}

\item {\em Source A:}  This galaxy (MCG+10-22-010) exhibits a heliocentric
recession velocity $781 \pm 80$ \kms\ \citep{fal1999} consistent with the
recession velocity of NGC 5907 and a Sloan $g'$ magnitude $g' = 14.987 \pm
0.003$ \citep{ade2011} and is morphologically classified as a dwarf irregular
galaxy \citep{ann2015}.  At the distance of NGC 5907, the absolute magnitude of
source A is $\approx -16.2$, i.e.\ roughly comparable to that of the SMC.  The
projected impact parameter of source A to the center of NGC 5907 is $\approx
117$ kpc and to the plane of the disk is $\approx 99$ kpc.

\item {\em Source C:}  This galaxy (LEDA 54419) exhibits a heliocentric
recession velocity $710$ \kms\ \citep{wen2000} consistent with the recession
velocity of NGC 5907 and a Sloan $g'$ magnitude $g' = 16.206 \pm 0.004$
\citep{wen2000} and is morphologically classified as a Magellanic irregular
galaxy \citep{ann2015}.  At the distance of NGC 5907, the absolute magnitude of
source C is $\approx -14.9$, i.e.\ roughly 0.3 times that of the SMC.  The
projected impact parameter of source C to the center of NGC 5907 is $\approx
56$ kpc and to the plane of the disk is $\approx 21$ kpc.

\item {\em Source D:}  This galaxy (2MASX J15140431+5630186) exhibits a Gaia
$G$ magnitude $G = 19.615 \pm 0.008$ \citep{gai2022}.  If it is at the distance
of NGC 5907, then the absolute magnitude of source D is $\approx -11.5$, i.e.\
roughly $1\%$ that of the SMC, and the projected impact parameter to the center
of NGC 5907 is $\approx 89$ kpc and to the plane of the disk is $\approx 43$
kpc.  This galaxy is of particular interest because it is located at the very
terminus of the western stream (and at the starting point of the putative
extension of the western stream), near the location of the thickening or
enhancement of the western stream noted in \S\ 5.2.  Because the galaxy is
included into the Gaia DR3 catalog, our analysis described in \S\ 4 attempted
(unsuccessfully, because it is not a point source) to model and subtract it.

\item {\em Source E:}  This galaxy (LEDA 2535522) exhibits a Gaia $G$ magnitude
$G = 20.44 \pm 0.01$ \citep{gai2022}.  If it is at the distance of NGC 5907,
then the absolute magnitude of source E is $\approx -10.7$, i.e.\ roughly
$0.6\%$ that of the SMC, and the projected impact parameter to the center of
NGC 5907 is $\approx 106$ kpc and to the plane of the disk is $\approx 102$
kpc.  As with source D, this galaxy is included into the Gaia DR3 catalog, and
our analysis attempted to model and subtract it.

\item {\em Source G:}  This galaxy (LEDA 2523331) exhibits a Gaia $G$ magnitude
$G = 20.58$ \citep{gai2022}.  If it is at the distance of NGC 5907, then the
absolute magnitude of source G is $\approx -10.6$, i.e.\ roughly $0.5\%$ that
of the SMC, and the projected impact parameter to the center of NGC 5907 is
$\approx 96$ kpc, and it is roughly in the plane of the disk.  As with source
D, this galaxy is included into the Gaia DR3 catalog, and our analysis
attempted to model and subtract it.

\end{itemize}
Properties of these galaxies are summarized in Table 3, which for each galaxy
lists the source, name, ICRS coordinates, heliocentric recession velocity
$v_{\rm rec}$, Sloan $g'$ or Gaia $G$ magnitude, morphological type, absolute
Sloan $g'$ or Gaia $G$ magnitude $M$, impact parameter $b$, and impact
parameter to the plane of the disk $b_{\rm disk}$.

Source F in Figure 12 might appear at first glance to be a dwarf galaxy in
close proximity to the eastern stream.  But our images resolve this ``source''
into a number discrete sources, which we interpret as a background galaxy group
or cluster.  There are several other background galaxy groups or clusters also
evident in the images.

We conclude that there are at least several (and possibly many more) dwarf
galaxies associated with NGC 5907 that may play roles as progenitor galaxies.
The few galaxies considered here are far from a complete inventory of dwarf
galaxies and possible dwarf galaxies associated with NGC 5907, and as is
discussed in \S\ 4, our processed images exhibit a large number of faint
sources, some fraction of which could be dwarf galaxies.  Multi-object
spectroscopy or multi-band imaging of faint sources in the field surrounding
NGC 5907 will be necessary to identify other dwarf galaxies associated with NGC
5907.

\subsection{Possible Confusion with Galactic Cirrus}

To assess possible confusion with Galactic cirrus in the direction of NGC 5907,
we examined (1) AKARI far-infrared all-sky survey maps at 65, 90, 140, and 165
$\mu$m \citep{doi2015} and (2) an interstellar reddening map derived from H~I
emission \citep{len2017}.  We found that at the Galactic coordinates $l =
91.58$ deg and $b = +51.09$ deg of the galaxy, there is negligible infrared
emission at any AKARI bandpass, and there is negligible interstellar reddening.
We therefore consider it highly unlikely that any of the very
low-surface-brightness features in the direction of the galaxy arise due to
Galactic cirrus.

\section{Summary and Discussion}

The results described in \S\ 5 confirm the overall picture of the galaxy NGC
5907 and its stellar stream advanced by \citet{van2019}:  the stellar stream
consists of a single curved structure that stretches $220$ kpc from the
brighter eastern stream, across the southern edge of the galaxy, to a fainter
western stream that bends to the north and then curls back south toward the
galaxy.  But these results also demonstrate that the the situation is more
subtle and complex in several respects:  (1) the western stream appears to
bifurcate near its apex, (2) there is an apparent gap of $\approx 6$ kpc in the
western stream due east of the galaxy, (3) there is no evidence of the remnant
of a progenitor galaxy within the eastern stream, although (4) there are many
other possible progenitor galaxies, including some that are quite close and at
least one that is located within the western stream, (5) there is another
structure that stretches $240$ kpc and that contains two very large, very
low-surface-brightness patches of emission, one of which was noted by
\citet{van2019} and another of which was not, and (6) there are other notable
new features, including a western ``horn,'' a southern ``spur,'' and a western
``hook.''

We consider several aspects of these results to be particularly significant as
follows:

First, we note that two different $N$-body simulations (i.e.\ by
\citealt{mar2010} and by \citealt{van2019}) predict two very different
configurations for the stellar stream, both of which apparently run counter to
observation (in one case with respect to the second loop and in the other case
with respect to the remnant of a nearly disrupted progenitor galaxy).  This
suggests to us that the boundary conditions of both simulations are very
significantly under constrained.  We propose that a correct and complete
understanding of the nature and origin of the stellar stream can be obtained
using $N$-body simulations only if additional boundary conditions can be
supplied, most crucially relating to the eastern stream and of source D.

Second, we are intrigued by the number and variety of stellar streams in the
vicinity of NGC 5907, including the eastern stream, the western stream, the
structure containing the linear feature and two patches of emission, and
possibly the western hook.  Given that more than 100 stellar streams are known
in the vicinity of the Milky Way \citep[e.g.][]{mat2022}, there is every reason
to suspect that similar networks of stellar streams might be found around other
galaxies, including NGC 5907.

Third, we are struck by the apparent gap in the western stream.  Gaps in
stellar streams may be caused by the impacts of dark ``subhalos'' or satellites
orbiting within the halos of massive galaxies \citep[e.g.][]{hel2016, kop2020}.
Hence the apparent gap in the western stream may be indicative of a dark
subhalo or satellite in the vicinity of the galaxy.  Further observations and
analysis are clearly required to confirm and interpret the apparent gap.

Finally, we are puzzled by the nature of the two very large, very
low-surface-brightness patches of emission.  If these patches are considered to
be galaxies, then they are extremely low-surface-brightness galaxies; if these
patches are not considered to be galaxies, then is not clear what they are, and
they presumably represent some new phenomenon with no known analog.  We
speculate that the presence of the patches is in some way related to the
presence of the tidal stream, although the morphology of the linear feature and
the patches together is vaguely reminiscent of ``jellyfish'' galaxies
\citep[e.g.][]{mor2018} or of the young, isolated stellar systems found in the
Virgo cluster \citep{jon2022}, both of which may be formed via ram-pressure
stripping of gas from a parent galaxy.

There is clearly more to be learned about the galaxy NGC 5907 and its stellar
streams, and we anticipate using Condor to obtain additional deep observations
of NGC 5907 and the NGC 5866 Group through its complement of broad- and
narrow-band filters.

\section*{Acknowledgments}

This material is based upon work supported by the National Science Foundation
under Grants 1910001, 2107954, and 2108234.  We gratefully acknowledge Chris
Mihos and an anonymous referee for very valuable comments on earlier drafts of
the manuscript; the staff of Dark Sky New Mexico, including Diana Hensley,
Michael Hensley, and the late Dennis Recla for their outstanding logistical and
technical support; and Yuri Petrunin for crafting six superb instruments.  This
work made use of the following software: astroalign \citep{ber2020}, astropy
\citep{2013A&A...558A..33A,2018AJ....156..123A}, django \citep{django}, Docker
\citep{mer2014a}, DrizzlePac \citep{gon2012}, NoiseChisel \citep{akh2015,
akh2019}, numba \citep{lam2015}, numpy \citep{har2020}, photutils
\citep{bra2020}, scipy \citep{vir2020}, SExtractor \citep{1996A&AS..117..393B}.

\section*{Data Availability}
 
All raw Condor data are available following an 18-month proprietary period.
All raw and processed data described here, including the coadded images of
Figures 1 through 6 and the mosaic image of Figure 7, are available on the
Condor web site https://condorarraytelescope.org/data\_access/ or by contacting
the corresponding author.

\bibliographystyle{mnras}
\bibliography{manuscript}

\bsp	% typesetting comment
\label{lastpage}
\end{document}